\documentclass[showpacs, twocolumn, preprintnumbers,amsmath,amssymb,amsfonts,amsthm]{revtex4}
\usepackage{amsthm, epsfig}
\usepackage{dcolumn, natbib}
\usepackage{bm}

\newtheoremstyle{note}
  {3pt}
  {3pt}
  {}
  {}
  {\itshape}
  {:}
  {.5em}
  {}

\theoremstyle{note}

\newtheoremstyle{citing}
  {3pt}
  {3pt}
  {\itshape}
  {}
  {\bfseries}
  {.}
  {.5em}
  {\thmnote{#3}}

\theoremstyle{citing}

\newtheoremstyle{break}
  {9pt}
  {9pt}
  {\itshape}
  {}
  {\bfseries}
  {.}
  {\newline}
  {}

\newcommand{\pr}{{\mathcal P}}

\let\lvert=|\let\rvert=|

\begin{document}


\title{Representation of protein structures based on frequency distributions of oriented cycles in contact graphs}

\author{A. M.~Lisewski and O.~Lichtarge}
\affiliation{
Department of Molecular and Human Genetics, Baylor College of Medicine, One Baylor Plaza, Houston, TX 77030, USA
}%


\begin{abstract}
We present a statistical approach to protein structure by introducing a representation of  protein folds based on simple observables defined as frequencies of oriented cycles in contact graphs. Motivated by the idea that these cycles may form a code of the entire protein structure we investigate its Shannon entropy. The latter surprisingly shows a characteristic transitory behavior when traced over different values of the geometric threshold $t$ which defines protein residues in contact. To account for this observation, we  propose a non-linear mechanical model--- formulated as a Hamiltonian system in which $t$ is regarded as a continuous parameter--- that allows to identify the behavior of the Shannon entropy as a first-order phase transition between a disordered and an ordered phase of the proposed mechanical system. The transition itself reflects the formation of protein structure when represented in terms of contact graph cycles, and it is identified as an example of a fragmentation transition known from several other statistical systems without a proper thermodynamic limit. Some interesting implications follow from our model including chirality, broken $t$-symmetry and one-dimensionality. Although defined from purely structural considerations, we further show that for native protein structures these observables follow some of the quantitative rules that are typically valid for amino acid types in protein sequences. Moreover, this relation between polypeptide structures and their amino acid sequences suggests a specific protein design alphabet.
\end{abstract}

\maketitle

{\section{Introduction}}
How should we represent a protein? Depending on the given scientific context there are several valid representations to choose from. For example, a protein may be viewed as a single point in phylogenetic tree spaces; or as a set of many (quantum) point particles interacting through Coulomb forces in 3D space; another legitimate representation is a finite sequence from the amino acid alphabet; and we may also describe a protein as a so-called contact graph, i.e. a graph defined through contacts between its amino acid residues. Indeed, the latter representation has recently provided insights into the structure of protein folds. For example, the reconstruction of the protein's native structure from the knowledge of the contact map's principal eigenvector \cite{poba2004}, or the idea that those graphs are {\it small-world networks} in which residues of high connectivity (``hubs'') play a significant role in the folding process \citep{vedo2002}. These and related studies indicate that contact graphs are a phenomenologically rich way of representing proteins. 

In the present work we introduce a representation of proteins derived from simple topological properties of contact graphs to demonstrate several aspects of protein folds. This representation is given through a family of observables which are the relative frequencies $q_k$ of oriented cycles of different lengths $k$ present in contact graphs. Every contact graph is composed from cycles of various lengths $k$, and so every $q_k$ simply is the frequency of ordered contact pairs of residues that are $(k-1)$ sequence steps apart. The frequency values directly depend on the choice of the geometrical threshold $t$ which defines contacts between protein residues (i.e., two amino acid residues are said to be in contact if their pairwise distance is smaller than $t$): the larger the threshold $t$ the more contacts we count in each $q_k$.  To investigate this dependence from an information theoretic point of view we introduce the Shannon entropy $H_c$ for the entire distribution of frequencies $\{ q_k \}$. Our first result is that for native protein structures $H_c$ is a non-monotonic and partially convex function of $t$. Interestingly, a characteristic minimum of the entropy at $t_*\approx 5.6 \AA$ localizes the convex region. This alone is remarkable for two reasons. First, it readily establishes a structural characteristic of proteins, since a local entropy minimum does not appear for totally random chains. For example, we show that for 3D Markovian random walks $H_c(t)$ is a monotonically increasing and linear function. Second, convexity in entropy functions for certain statistical systems is known to be an indicator for {\it phase transitions} \citep{gr2001a, chgo2001}. However, at this point we only have the entropy function $H_c$ at hand, raising the important question whether we can identify any actual system that undergoes a phase transition indicated by convex entropy regions.

To address this question we introduce a mechanical model, defined as a Hamiltonian system in which $t$ plays the role of a time parameter, that accounts for the functional behavior of the entropy function. This model follows naturally from the Shannon entropy $H_c$ and it implies several intriguing aspects from non-linear Hamiltonian mechanics (instability, broken time reversal symmetry, chirality) and from non-equilibrium statistical mechanics (first order phase transitions). In fact, we show that this Hamiltonian system can be interpreted as a rather simple classical mechanical system of many non-linearly coupled constituents (which we call mechanical particles) moving in one dimension. We demonstrate that the Hamilton function of our system is a sum of a non-linear kinetic term $K$ and a hypothetical potential energy term $U$, which means that we do not further specify the functional form of $U$. This, however, does not invalidate the arguments given in this work because the relevant points of our analysis solely depend on the well defined kinetic term $K$. Using this formulation, we can regard convex regions in the entropy function as locations of phase transitions where the mechanical system becomes inhomogeneous as several of its particles coalesce into fragments when a certain value range of the geometrical parameter $t$ has been reached. We argue that this transition and its concurrent fragmentation mainly reflect the formation of protein secondary structure when going from small to larger values of $t$.  Hereby, we also show that there occur two adjacent first-order phase transitions. They determine a special value of the geometrical threshold parameter which turns out to be consistent with previous studies on contact graphs \citep{hile1992, veku1997}. Contact graphs defined upon the choice of this numerical value are known to carry the entire structural information \citep{veku1997}. Until now, its choice was motivated mainly by heuristic arguments and our analysis provides physical arguments instead.\smallskip

Transitions of the kind reported here are sometimes called (multi-)fragmentation transitions, and so far this phenomenon has been discussed for physical systems far from the thermodynamic limit such as atomic nuclei, mesoscopic systems and self-gravitating objects \citep{grma1997, gross1997, chgo2001, gr2003}. With the present work, a fragmentation transition is proposed in the protein context for the first time. In this manner we develop a framework where the formation of protein structure can be studied as a phase transition of a classical mechanical model that literally {\it evolves} in the threshold parameter $t$. This is a new alternative to the traditional view on proteins as multi-particle quantum physical systems that evolve in physical time.  Thus we suggest a new statistical and mechanical representation of protein structures.

To substantiate the significance of these results, we argue that our proposed representation of proteins is a specific candidate for the {\it protein design alphabet} \citep{dokh2004}. In doing so, we discover some striking similarities between the usage of amino acid types in protein sequences and the usage of the purely structurally defined family of cycle lengths $k$ in proteins. For instance, we show that at most twenty different cycle lengths are necessary to encode the structure of a contact graph when defined upon the distinguished minimum value of the contact threshold, $t = t_*$. Using the same threshold value, we further reveal a strong linear correlation between protein sequence length and the resulting number of contact graph cycles. Interestingly, this correlation sharply drops for smaller values of the geometrical contact threshold making the minimum of the entropy $H_c(t)$ peculiar again. These similarities establish a relation between protein sequence and structure and they strongly support evidence that amino acid sequences contain structural information encoded in lengths of contact graph cycles.\smallskip

This paper is organized as follows. In the next section we define the statistical observables in question and introduce their frequency distribution as well as the associated Shannon entropy $H_c$. In Section \ref{obs} we describe and discuss the functional behavior of $H_c(t)$ by considering an ensemble of many native protein structures obtained from the DALI/FSSP database. Here, we also elaborate the idea of looking at our representation as a protein design alphabet, and discuss implications to the protein design problem. Section \ref{hamil} introduces the mechanical model-- in terms of a Hamiltonian system-- for the cycle frequencies $q_k$. In this section several properties and implications of our model formulation are analyzed. They include instability, broken $t$-symmetry (Section \ref{eqmotion}), chirality (Section \ref{cst}), and first-order phase transitions (Section \ref{fopt}). We close with section \ref{conclusion} where we draw several conclusions and briefly outline possible future directions of our work.\medskip

\section{Contact graphs, oriented cycles and their distribution\label{distribution}}

Our model observables are frequencies of oriented contact graph cycles and they form a discrete probability distribution when taking an ensemble average over a statistical ensemble $\pr$ of many protein structures. Every ensemble member with an amino acid sequence of integer length $L$ is identified by its contact matrix $C \in \pr$. The protein residues are represented by the ordered set $\{1,\ldots, L\}$ such that the integer $1$ is associated with the N terminal residue and $L$ is identified with the C terminal. The contact matrix is derived from the spatial coordinates of the carbon C$_\alpha$ atoms taken from the protein's native structure. Since each C$_\alpha$ atom uniquely corresponds to a single residue of the amino acid chain, one can decide whether or not any two arbitrary residues are close to each other by using a Euclidean distance threshold $t$ of typically around $ 9 \AA$ \citep{veku1997} (In what follows, we will always represent $t$ as a dimensionless multiplier of $1 \AA = 10^{-10}$ m.). A contact matrix is a $L \times L$ symmetric matrix with zero entries for all residue pairs whose mutual distance is larger than $t$, and with entries equal to one for all remaining pairs which are called contact pairs. Since in a protein sequence every nearest and next to nearest residues are in contact, we put $C(i,j) = 1$ for all $i,j \in \{1, \ldots, L\}$ with $|i-j| \leq 2$.\smallskip

 To define these observables,  we first ask for the number $q_k$ of contact pairs $(i,j)$ for any given sequence distance $ k-1 = j - i$ with $2 \leq k -1 \leq L-1$ and with $j > i$. We obtain this number by evaluating a conditional sum over the entries in the contact matrix $C$, i.e.
\begin{equation}
\label{pk1}
q_k = \sum_{\substack{j>i,\\ k - 1 = j - i}} C(i,j) \,.
\end{equation}
This number can be put readily into a graph theoretic context. Since any contact matrix represents a graph $G$ of $L$ nodes and with arcs $(i,j)$ between exactly those nodes which are in contact, the frequency $q_k$ gives the number of closed paths of length $k$ which are followed in their first $k-1$ steps according to the direction imposed by the ordered set of the residues $\{1, 2, \ldots , L\}$. Thus it is the ordering of the protein's amino acid chain going from the N to the C terminal that gives rise to a N-C direction all closed paths counted by $q_k$. In this way, we discard all cycles (i.e., closed paths) in $G$ that have more than {\it one} step opposite to the N-C direction. 

Next we give a statistical meaning to each $q_k$ by taking the ensemble average over all contact matrices in $\pr$. Let us consider the number $L_m$ being the maximum amino acid sequence length in $\pr$. For each $3 \leq k \leq L_m$ we calculate the relative ensemble frequency 
$
\bar q_k = N^{-1} \,\sum_{C \in \pr} q_k  
$
with $N = \sum_k \sum_{C \in \pr} q_k$, and thus we obtain a discrete probability distribution $\{\bar q_k \} \equiv \{\bar q_3, \ldots, \bar q_{L_m}\}$ (Each $\bar q_k$ is henceforth denoted as $q_k$.). We regard this distribution as a statistical distribution among the letters of an alphabet consisting of all possible N-C cycle lengths $3 \leq k \leq L_m$. It is convenient to introduce the index set valued function $s(\pr)$ such that $k \in s(\pr)$ iff $q_k > 0$, and to denote $n^+ = |s(\pr)|$. We can now define the Shannon entropy $H_c$ for $\{q_k\}$ as
\begin{equation}
H_c = -\sum_{k \in s(\pr)} q_k \log q_k \,.
\end{equation} 
Here, the individual frequencies $q_k$ are functions of the distance cut-off $t$, because each contact matrix $ C \in \pr$ is defined upon its choice. Therefore, $n^+$ and $H_c$ are both functions of $t$ and our next aim is to explore their functional behavior.

\section{Functional behavior of the entropy $H_c$ and the protein design alphabet\label{obs}}

To study $H_c(t)$ for an ensemble of real proteins, we have realized $\pr$ using the DALI/FSSP database of native protein structures \citep{fssp} (We have considered only the first available chain/domain in each database file and the maximum domain length in this database determines $L_m$, which turns out to be $L_m=1058$.). Fig. \ref{fig1} shows the resulting entropy $H_c$ as a function of the distance threshold $t$.  We see an onset and an expected increase of $H_c$ at approximately $t \approx 3$. However, the monotonic concave increase of $H_c$ is interrupted at a range $4.8 \lesssim t \lesssim 5.8$, where $H_c$ decreases between a maximum at $t^* \approx 4.8$ and a minimum at $t_* \approx 5.6$. The local minimum also marks the region where $H_c$ is not a concave function of $t$, but where it is instead characterized by convexity. The drop of $H_c$ to a local minimum at $t_*$ shows that most of N-C oriented cycles in $\pr$ can be effectively encoded with an alphabet of only $\bar n_* = \exp[H_c(t_*)]$ letters. This statement follows since $n^+(t_*) \approx 670$, while the function $n^+$ saturates at a comparable value $n^+ \approx 920$ for large $t$.  And at local minimum with $H_c(t_*) \approx 2.97$ we get $\bar n_* \approx \exp(2.97) = 19.5$, a much smaller number than the total number of realized letters of the cycle alphabet, $n^+(t_*) \approx 670$. 

To further confirm these findings we tested different choices of the ensemble $\pr$ by lowering the maximum protein length $L_m$ of chosen protein structures from the DALI/FSSP database. All choices show the local maximum and the minimum at the same values of $t^*$ and $t_*$. However, the heights of the extrema change-- in particular, $H_c(t_*)$ as a function of $L_m$ is monotonically increasing with an asymptotic value of $\bar n_* \approx 19.5$ for values of $L_m$ larger than $\sim\!600$. 

For comparison, Fig. \ref{fig1} also includes the entropy function $H_c(t)$ but for an ensemble ${\cal P}_r$ of 3D Markovian random walks with the same total number of randomly generated chains, i.e. $|{\cal P}_r| = |{\cal P}|$, and with the same individual chain lengths as given by the DALI/FSSP ensemble of structures. Each chain in ${\cal P}_r$ has a fixed step length of $\Delta s = 3.8 \AA$ which is comparable to the average distance between two neighboring C$_\alpha$ atoms in proteins. In contrast to the previous case, the entropy function does not show any convex intruder (in fact, it is almost a linear function of $t$ within the $t$-range considered), and so we conclude that the local minimum along with the convex region to which it corresponds is a characteristic signature of native proteins.  

One explanation for the local minimum of $H_c$  could be that most of the oriented cycles in the distribution $\{q_k\}$ correspond to secondary structures forming alpha-helices. This is supported by the fact that each $2 \pi$ turn in an alpha-coil typically results in three N-C oriented loops of length 3, 4, and 5 (they correspond, respectively, to residues that are in contact and which are 2, 3 and 4 steps apart in the amino acid chain). Since alpha-helices have an average period of $3.6$ chain steps and because it is known that alpha-helices make up a fraction of around 30$\%$ of the total secondary structure found in proteins, a large portion of the cycle distribution $\{q_k\}$ should represent short loops of length 3, 4, or 5.  And we indeed verify that the relative frequencies $q_3$, $q_4$, and $q_5$ contribute most in $\{q_k\}$. However, we think that this qualitative explanation cannot account for the characterstic values of the entropy function found at the local minimum, as we find that $\bar n_*\approx 19.5$ is larger than 3 = $| \{3,4,5\}|$, and hence a three letter cycle alphabet of lenghts $\{3,4,5\}$ cannot encode the information given in $\{q_k\}$.  

Alternatively, one sees a connection of the asymptotic value of $\bar n_* \approx 19.5$ and the twenty amino acids types that form the alphabet of all protein sequences---raising the possibility that there is a deeper understanding behind the similarity between the two numbers. Moreover, the total number of twenty alphabet letters turns out to be special even when looking at the distribution $\{q_k\}$ itself. Figure \ref{fig2} shows two examples (for $t = 5.0$ and for $t = 9.0$) of this distribution extracted again from the DALI/FSSP database with maximum available chain length $L_m = 1058$. The resulting graphs are partitioned into three different ranges where each range has a different scaling behavior: the larger the cycle length $k$ the steeper the descent of the graph. Interestingly, the first range A is characterized by a rather shallow decrease in absolute frequencies $q_k$. In contrast, range B for $k \in \{23,\ldots, \sim\!400\}$ is clearly different from A by its steeper slope ($m_B \approx -1.7$ in double-logarithmic scale). And, surprisingly, the boundary point between range A and range B is located almost exactly after the the first twenty N-C cycle lengths $\{3,\ldots, 22\}$. Since the cycle frequencies in range A have by far the highest values within the entire distribution, we again have a situation where an effective set of twenty letters emerges from our underlying alphabet of cycles. If not given by pure chance, these correspondences are remarkable  because at no point have we included any information from the amino acid sequence into our study of the cycle length distributions; we recall that the latter are based only on the structure of contact graphs. Are these correspondences telling us that amino acid sequences and cycle lengths in protein contact graphs have something fundamental in common?
 
To address this question, we argue that contact graph cycle lengths are a candidate for the hypothetical protein alphabet as discussed recently by Dokholyan \citep{dokh2004}. According to his observations, less than statistically expected amino acid types are used in natural protein structures, from which he concludes that amino acids might not be the natural code to describe protein structure. One consequence that he draws from this fact is that a family of more than twenty ``abstract objects'' \citep{dokh2004} constitutes this natural coding alphabet for structures. Identifying this structural alphabet would be an important step forward in the {\it protein design problem}---which is the difficult task of designing a sequence that folds into a targeted polypeptide structure.  Using our structural representation, our claim is that these abstract objects are N-C oriented cycles. 

To support this claim, we first look at the {\it amino acid usage} given a mean protein domain length obtained by Dokholyan \citep{dokh2004} from the DALI/FSSP database (Fig. \ref{fig3}). The resulting graph shows how many different amino acid types are used given an average protein domain length; it is characterized by a strong increase from 10 to about 18 amino acid types when going to domain lengths up to  $\sim\!150$ amino acids, followed by a saturation at 20 amino acid types for longer domains. Now returning to a representation based on cycle lengths,  we observe in Fig. \ref{fig3} an almost identical functional behavior (within the uncertainty given by the error bars) for the effective cycle number, $\bar n_*$, evaluated in the same range of mean domain lengths taken from DALI/FSSP. This surprisingly shows that amino acid usage in sequences and cycle length usage in structures with corresponding sequence lengths behave the same way. In a related analysis, shown in Fig. \ref{fig4}, each sequence length from of DALI/FSSP is plotted against its corresponding total number of cycles evaluated at $t = t_*$. This graph shows a well exposed linear dependence between both numbers with a optimum linear slope of $m \approx 0.91$. Also in Fig. \ref{fig4} are shown the resulting Pearson correlation coefficient $r$ and the linear regression slope $m$ as functions of $t$. We observe that $r$ strongly rises from $r \approx 0.1$ (uncorrelated state) to $r \approx 1$ (strongly correlated state) between $t\approx 4$ and $t\approx t_*$, indicating that at $t \approx t_*$ correlation between sequence length and cycle number reaches its absolute maximum (upper left insert in Fig. \ref{fig4}). At the same time, interestingly, the optimum linear slope $m$ approaches a value of one in the neighborhood of $t\approx t_*$ (lower right insert in Fig. \ref{fig4}). It is therefore evident that, in a statistical sense, the length of an amino acid sequence and the number of corresponding contact graph cycles are the same when evaluated at the characteristic minimum $t_*$ of the entropy function $H_c(t)$. In this way we have detected two quantitative relations between sequence and structure. Both relations suggest that the family of all possible cycle lengths $k$ constitutes a protein design alphabet, thus directly supporting Dokholyan's original conjecture.\smallskip 

Taking into account these results we want to briefly discuss their implication for the protein design problem. Suppose we want to design a certain protein, or any protein-like amino acid polypeptide, and thus we set up its target structure by specifying the 3D coordinates of its C$_\alpha$ atoms which together form the polypeptide's carbon backbone of length $L$. What constraints does this information already put on the amino acid sequence $S_{aa}$ that would ideally fold into the target?  Given the initial data we first may derive the structure of the corresponding contact graph evaluated at minimum entropy contact threshold $t_* \approx 5.6$. This graph concurrently defines a sequence $S$ of N-C oriented cycles with lengths $3 \leq k \leq L$ or, equivalently, an equally long sequence $S'$ of ordered contact pairs of residues being $2 \leq (k-1) \leq (L-1)$ sequence steps apart. Thus $S$ lists the cycle lengths of our target structure, e.g. $S = \{3, 4, 29, 3,\ldots\}$ and $S' =\{2, 3, 28, 2,\ldots\}$. At this point we know that we can safely reduce the number of different cycle lengths in $S$ to a maximum of twenty different lengths (see Fig. \ref{fig3}). We stress, though, that we do not know a priori {\it which} cycle lengths can be discarded from the original alphabet. Although we have shown that statistically the twenty shortest cycle lengths contribute most to the distribution $\{q_k\}$, this does not imply that all of them are non-redundant, i.e. not all of them have to be necessarily indispensable to represent the structure. So the first difficulty is the identification of the twenty or less cycle alphabet letters which are necessary to encode the structure of the contact graph. Our results further indicate that the length of the amino acid sequence, $|S_{aa}| = L$, and the total number of cycles in $S$ should coincide (see Fig. \ref{fig4}), i.e. $|S| \approx L.$ Thus we expect $S_{aa}$ to have both the same length and the same number of different letters as $S$, which suggests that there exists a one-to-one correspondence between both sequences. But here we come to a second obstacle: we do not know the natural order of the code letters in $S$. In other words, we need to know what given position in $S$ corresponds to what position in $S_{aa}$ so that translation in the correct order of letters would eventually generate the aimed fold. Hence we conclude that although we have revealed a connection between the targeted structure and the sequence it takes for it to fold, protein design still remains a problematic task, of course.

\smallskip

\section{Model formulation: The non-conservative Hamiltonian flow\label{hamil}} 
\subsection{Equations of motion\label{eqmotion}}
{{We are now going to introduce a theoretical model which follows naturally from the entropy function $H_c(t)$, and which gives a novel physical description of its functional behavior. This model is cast in the framework of Hamiltonian mechanics in which the distance threshold $t$ takes the role of the usual time parameter. As we shall see in the following sections, this formulation allows us to interpret any convex region in $H_c(t)$ (see Fig. \ref{fig1}) as a location where a first order phase transition occurs in a multi-particle and non-linear Hamiltonian system evolving in $t$. We further show that this evolution is not $t$-symmetric and that the mechanical particles themselves carry chirality.}} 

To reach our goals we initially  consider each relative frequency $q_k$ as a continuous function of $t$. This is an approximation because each corresponding absolute number of N-C oriented $k$-cycles can increase only step-wise as $t$ increases. However, as long as the total number of measured cycles remains much larger than its (discontinuous) increase due to a small change in $t$ the continuum approximation should hold. The continuum condition is well met for large enough ensembles such as the DALI/FSSP database, and our remaining discussion is based upon its validity. 

At each local extremum of $H_c(t)$ it is 
\begin{equation}
\label{extremumcond}
\dot H_c(t) = 0 = - \sum_k \dot q_k(t) \, [1 + \log q_k(t) ] \,,
\end{equation}
with $t \in \{t^*, t_*\}$. Each term of this sum over $k$ may therefore be written as
\begin{equation}
\label{qp1}
- \dot q_k (1 + \log q_k) = p_k
\end{equation}
with $p_k \in {\mathbb R}$, and so condition (\ref{extremumcond}) reads as $\sum_k p_k = 0$. We now take a more general view on Eq. (\ref{qp1}) and put forward the assumption that it remains valid for all values of $t$, including the extremum points, where $n^+(t)$ is non-zero at the same time. As a consequence, the right-hand-side of the extremum condition $\sum_k p_k = 0$ may become positive or negative. In this situation Eq. (\ref{qp1})  is read as a proportionality statement between each ``particle velocity'' $\dot q_k$ and the corresponding ``momentum'' $p_k$, and so we may rewrite it as a Hamilton equation for each frequency $q_k$,  
\begin{eqnarray}
\label{hamil1}
\dot q_k  &=& \frac{\partial}{\partial p_k} (K+U) =  \frac{\partial}{\partial p_k} \left ( \sum_k  \frac{p_k^2}{2 m_k} + U \right )\,, 
\end{eqnarray}
where $K = \sum_k p_k^2/(2 m_k)$ is the total ''kinetic energy''. The proportionality factor between velocity $\dot q_k$ and momentum $p_k$ is the function 
\begin{equation}
m_k = -(1 + \log q_k)
\end{equation}
which we refer to as the  ``particle mass'' of the $k$th particle, and $U = U(q_3,\ldots, q_{L_m}; t)$ is a ``potential energy'' function. With the Hamilton function $(K + U)$ we also introduce the Hamilton equation for each momentum $p_k$,
\begin{equation}
\label{hamil2}
\dot p_k = -\frac{\partial}{\partial q_k} (K + U) = - \frac{1}{2 q_k} \left (\frac{p_k}{m_k} \right )^2 -\frac{\partial U}{\partial q_k} \,.
\end{equation}
Together with the normalization and the positivity constraints for the frequencies, $\sum_k q_k = \sum_k |q_k| = 1$, Eq. (\ref{hamil1}) and (\ref{hamil2}) constitute our hypothetical Hamiltonian flow in the phase space of the $L_m$ conjugate pairs $(q_k, p_k)$. Thus each pair $(q_k, p_k)$ is understood as a ``mechanical particle'' with mass $m_k$.  The system itself is hypothetical because we have not yet identified the potential energy function $U$. Note that normalization and positivity really are additional constraints here, because in our case Hamilton's equations are non-linear in all conjugate variable pairs. We also remark that the Hamiltonian system is inherently unstable: considering {\it free dynamics} by putting $U=0$ results in a negative change of the momentum, according to Eq. (\ref{hamil2}). This behavior marks a clear deviation from Newton's first law.

\subsection{Chirality, statistics, and $t$-symmetry\label{cst}} 

Let us look at a single particle with index $k$ in phase space. Hamilton's equations for $(q_k, p_k)$ imply a special case for those particles with ``positions'' $q_k$ near the $1/e$ point, because the mass $m_k$ vanishes at any $q_k = 1/e$. Thus if we demand that each particle velocity $\dot q_k$  should be finite, then its momentum $p_k$ must vanish sufficiently fast at the moment of crossing the $1/e$ point. There are four possible situations of how this crossing can occur.  Let $t_c$  be the crossing time point and let us observe the particle's behavior in some small neighborhood of $t_c$. In the first possibility (type I, see Fig. \ref{fig5}), $q_k$ approaches the $1/e$ point from below, that is from the interval $o = (0,\,1/e)$, it crosses the point once at $t = t_c$, and eventually it returns to region $o$. Concurrently, the particle's momentum $p_k$ is positive for all $t < t_c$, it vanishes at the crossing point, and finally changes its original sign. We introduce the particle's chirality $\chi_k(t)$ defined as
\begin{equation}
\chi_k = \frac{ \dot q_k \, p_k}{|\dot q_k| \, |p_k |} = \frac{m_k}{|m_k|}\,,
\end{equation}
and call a particle with $\chi_k = -1$ left-handed, while a particle  with $\chi_k = + 1$ we name right-handed. Clearly, for a type I crossing, $\chi_k(t)$ is positive for all $t < t_c$, and it remains so for $t > t_c$; thus velocity $\dot q_k$ and momentum $p_k$ are always parallel and no change in the particle's chiral state occurs. The situation is different for the other possible crossings. In fact, whenever it is $q_k > 1/e$, the particle's mass $m_k$ becomes negative and therefore the particle itself must be left-handed. Normalization of $\{q_k\}$ requires that there can be at most two left-handed particles. In this manner we find two crossing types (types III and IV in Fig. \ref{fig5}) where a change in chirality occurs (type III means right-handed at $t < t_c$, and type IV is initially left-handed at $t < t_c$), and two others (type I and type II in Fig. \ref{fig5}) where the chiral state is conserved (type I is right-handed for $t < t_c$, while type III is initially left-handed). Thus crossings of the $1/e$ line in the particle positions $q_k$ can change parity.

Given the above rules we want to derive {\it crossing statistics} for many particles. Let the $N$-tuple $(\chi_1, \ldots, \chi_N)$ represent the $N$-particle chiral state at times $t < t_c$ before crossing. What is the expected fraction $\nu^{(1)}_i(N)$ of right-handed particles after one crossing?  This fraction must be a function of the particle number $N$ and of the three possible initial states $ i \in \{0,1,2\}$ respectively corresponding to the system initially having zero, one, or two left-handed particles.  To obtain $\nu^{(1)}_i(N)$ we presume that each crossing type I-IV is equally probable, and solve the resulting combinatorial problem for all $N \geq 3$:
\begin{eqnarray}
\label{statistics}
\nonumber
\nu_0^{(1)}(N) &=& \frac{3N^2 - 2N + 1}{3N^2 + N}\,,\\
\nu_1^{(1)}(N) &=& \frac{3N^2 - 5N + 3}{3N^2 - N}\,,\\
\nonumber
\nu^{(1)}_2(N) &=& 1 - \frac{5}{4N} \,.
\end{eqnarray}
All three values monotonically approach 1 as for large $N$ the vast majority of the particles must be right-handed. On the other hand, for small $N$ we find a decrease of the mean fraction $\bar \nu^{(1)}$ reflecting the statistical preference of the system being mostly left-handed.  For instance, at $N=3$ the mean fraction becomes
$
\bar\nu^{(1)}(N=3) = \frac{1}{3} \sum_{i=0}^2 \nu_i^{(1)}(3) \approx 0.65\,,
$ 
for $N=2$ we get  $\nu^{(1)}(2) = 1/4$, and in the trivial case of $N=1$ we have $\bar \nu^{(1)}(1) = 0$.  Further studies should allow multiple crossings resulting in a calculation of $\bar\nu^{(c)}(N)$ for $c > 1$.\smallskip

Furthermore, we argue that the Hamiltonian system has {\it broken time reversal invariance}, i.e. it is a non-conservative Hamiltonian system. Inverting the sign of the velocities, $\dot q_k \rightarrow -\dot q_k$, in Eq. (\ref{hamil1}) accounts for a concurrent change in the sign of the momenta $p_k$, but time reversal symmetry cannot be sustained in the momentum Eq. (\ref{hamil2}) because the first term on the right hand side is negative for all values $q_k$ and $p_k$. \smallskip 

\subsection{First-order phase transitions\label{fopt}} 
The statistical physics of systems with a relatively small number of degrees of freedom has been recently investigated based on their micro-canonical properties \citep{gross1997, gr2001a}. These systems often cannot be described properly in a thermodynamic limit which is valid for large and locally homogeneous systems. Therefore they are referred to as {\it small} \citep{gross1997, vohi2002} or as {\it finite} \citep{chgo2001} statistical systems. Known examples of small systems are self-gravitating objects (e.g., stars) and atomic nuclei \cite{gr2003}, and our aim is to describe the behavior of our mechanical system within this framework.\smallskip 

\begin{figure}[t]
\epsfig{file=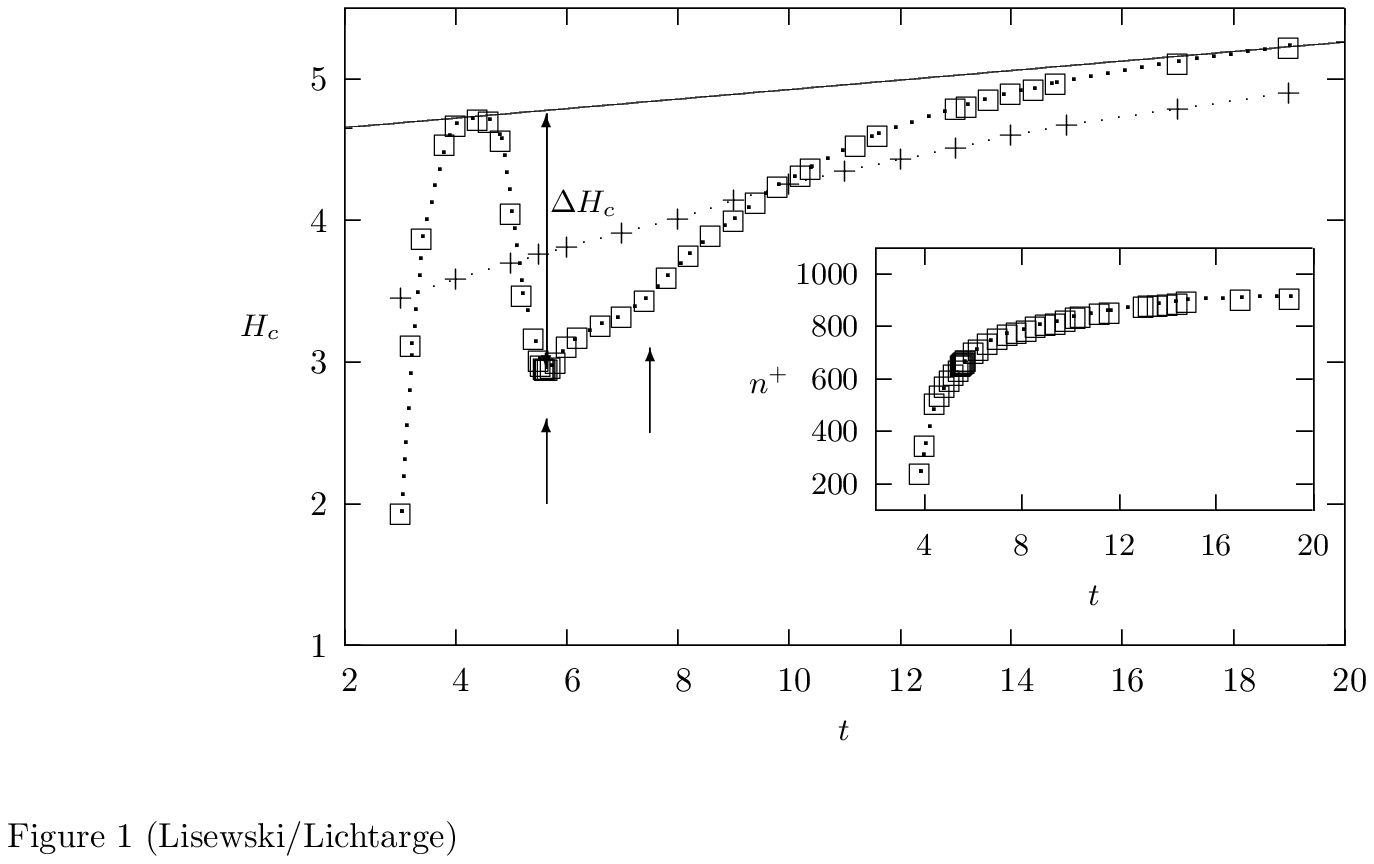}
\caption{The large graph shows the entropy $H_c$ as a function of $t$ (measured in $\AA$) derived from the DALI/FSSP database of protein structures. For comparison, points labeled with '+' give the  entropy for domains created by Markovian 3D walks with step width $\Delta s= 3.8 \AA$. The insert shows the number $n^+$ of observables $q_k$ with $q_k > 0$ for all $k \in \{3,\ldots,L_m\}$ in the same $t$-interval as $H_c$. The two small arrows pointing up depict centers of convex regions in $H_c$. $\Delta H_c$ indicates the depth of the first convex region estimated as the maximum vertical distance between the double tangent (the straight line above) and the local convex region of $H_c$. The double tangent gives the concave hull of $H_c(t)$ (Gibbs construction).\label{fig1}}
\end{figure}

\begin{figure}[t]
\epsfig{file=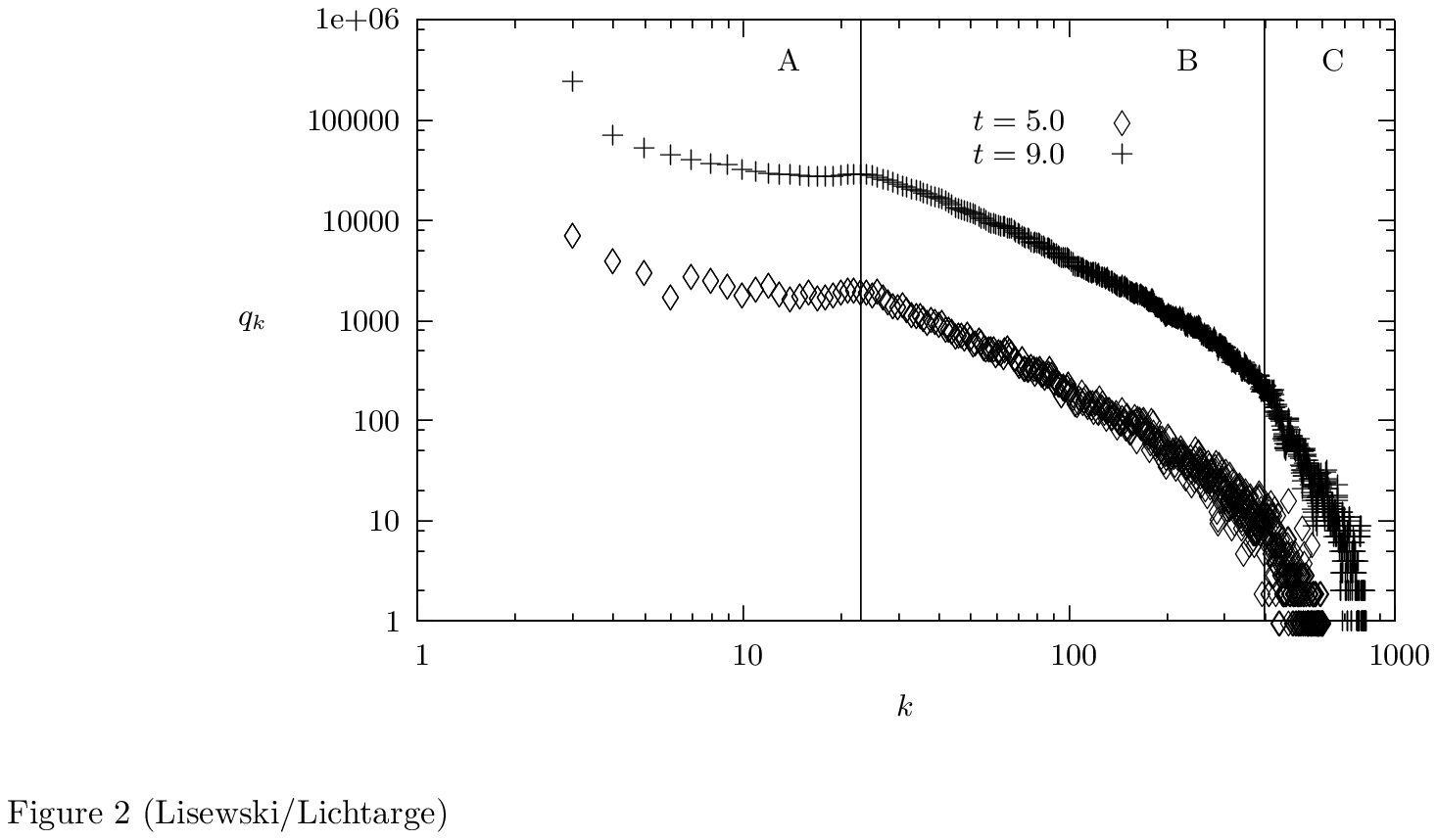}
\caption{The distribution $\{q_k\}$ of N-C cycle frequencies extracted from the DALI/FSSP database with maximum available chain length $L_m = 1058$. Both curves show three different scaling ranges, i.e. ranges with different slopes. The first vertical line at $k = 23$ indicates the boundary of the first twenty cycle lengths $k \in \{3,\ldots,22\}$; this line intersects with the boundary (``knee'') between the first (range A) and the second  scaling range (range B) The linear slope $m_B$ at range B is for both graphs $m_B \approx -1.7$, while for range C we have $m_C \approx -7.0$. Range A shows a clear deviation from in linearity in double logarithmic scale.\label{fig2}}
\end{figure}

\begin{figure}[t]
\epsfig{file=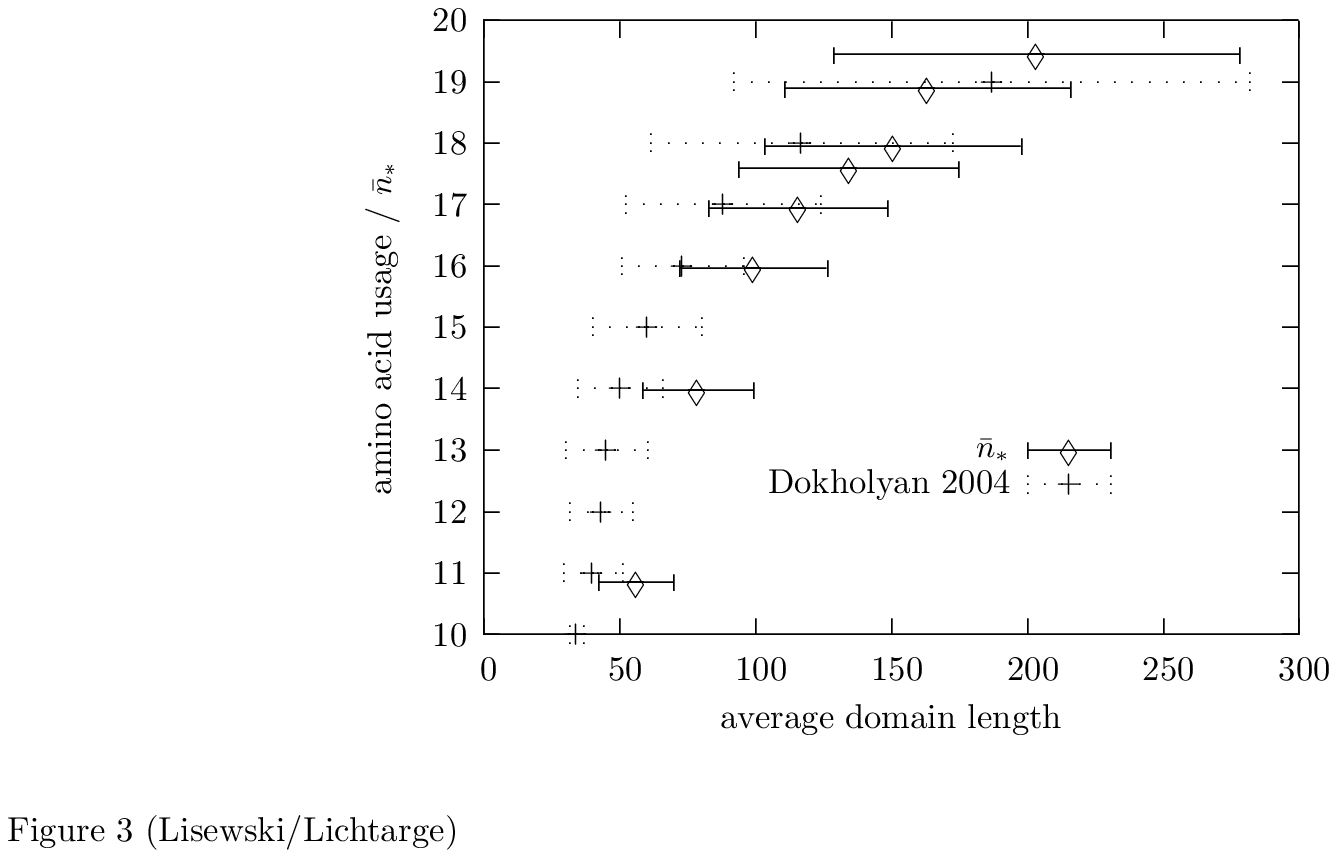}
\caption{Comparison of the amino acid usage (graph depicted with '+' and taken from \citep{dokh2004}) with the effective number $\bar n_*$ of oriented cycles (graph depicted with '$\diamond$') within a range of mean domain lengths of structures taken from DALI/FSSP. Standard deviations in domain length are included in both graphs.\label{fig3}}
\end{figure}

\begin{figure}[t]
\epsfig{file=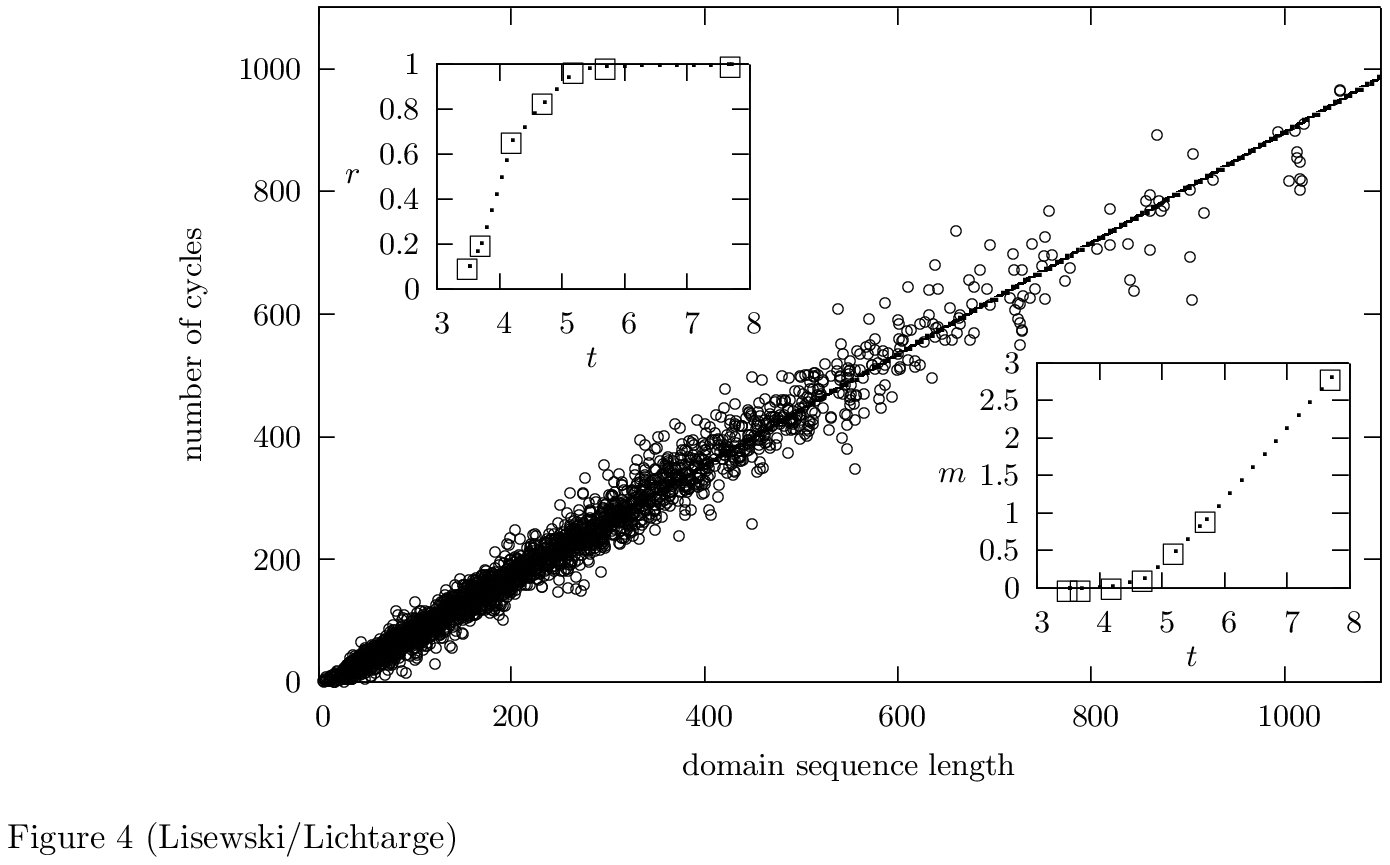}
\caption{Scatter plot of DALI/FSSP domain sequence lengths and their corresponding numbers of estimated oriented contact graph cycles evaluated at a minimum threshold value $t_*= 5.6 \AA$. Linear regression gives a slope of $0.91$ and an offset of $-9.5$. The upper-left insert shows the Pearson correlation coefficient as a function of $t$, while the lower-right insert does the same for the slop $m$ of the resulting dataset given by linear regression.\label{fig4}}
\end{figure}

\begin{figure}[t]
\epsfig{file=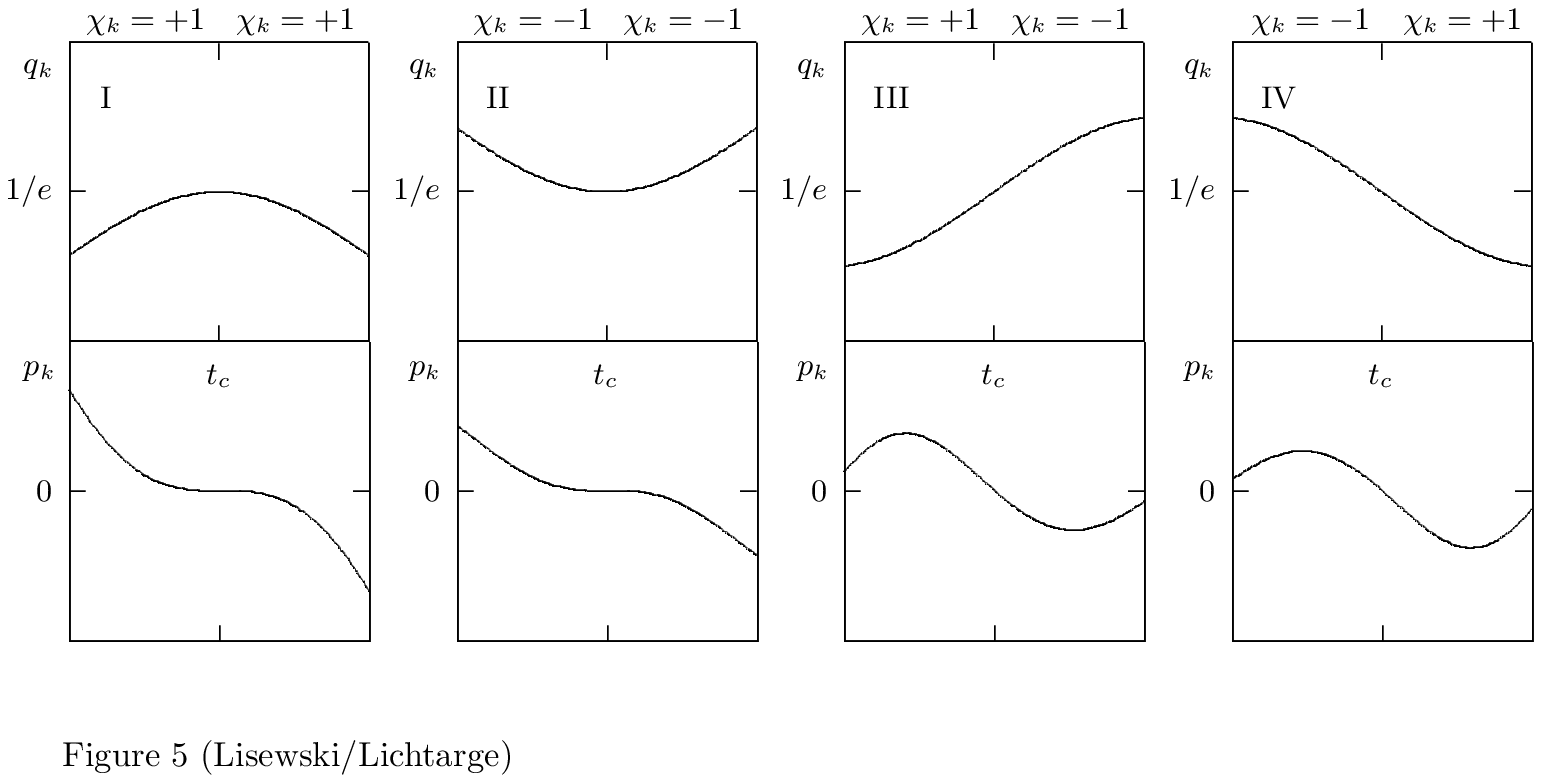}
\caption{Schematic illustration of the four crossing types of the $q_k =1/e$ point in phase space of $(q_k, p_k)$. Type I and II induce a change in chirality $\chi_k$ while III and IV do not. \label{fig5}}
\end{figure}

The entropy function is an indicator for phase transitions in small or finite statistical systems \citep{gross1997, gr2001b, chgo2001}. Phase transitions are then characterized in terms of the topology of the entropy function: concavity indicates a pure phase while convexity reflects a transitory behavior where several different phases coexist. If the entropy $S$ is a function of several observables $\{x_1,\ldots, x_n\}$ then the largest eigenvalue $\lambda_1$ of the curvature matrix with entries $
\partial_{x_l}\partial_{x_l} S$ with $(m,l) \in \{1,\ldots, n\}^2
$ 
determines the order of the phase transition: for $\lambda_1 > 0$ the transition is of first order while second order phase transitions prerequisite a vanishing $\lambda_1$. The eigenvector corresponding to the largest eigenvalue constitutes the local order parameter. In our case, the entropy function is the Shannon entropy $H_c$ and the observables are the absolute numbers of different N-C loop lengths for a given ensemble $\pr$. Since there is a one-to-one correspondence between the threshold values $t$ and the absolute numbers of the loop lengths within an statistical ensemble $ \pr$ (they are strictly monotonically increasing functions of $t$), it is sufficient to consider $t$ as the only independent observable in $H_c$. Hence convex regions with positive second derivative of $H_c(t)$ indicate a first order phase transition, and the local order parameter is
\begin{equation}
-\frac{d^2}{dt^2}\sum_k q_k \log q_k = \sum_k \dot p_k \,.
\end{equation}
There is a convex region of the entropy function around $t_*=5.6$ (Fig. \ref{fig1}) separating two concave regions at $3 \lesssim t \lesssim 5.4$ and at $t \gtrsim 6$. Therefore a first order phase transition occurs at $t_*$.  Interestingly, we see a second convex region of the entropy function in the interval $6.5 \lesssim t \lesssim 9$ being much less prominent than the first at $5.4 \lesssim t \lesssim 5.8$. This observation implies that there are two first order phase transitions occurring close to each other.

First order phase transitions in small systems signal the fragmentation of the system constituents into clusters, and thus indicate a preference of the system to become inhomogeneous and to separate into different phases. From this point of view we can interpret the  behavior of $H_c(t)$ as a transition from a random  phase to an ordered phase where the formation of secondary protein structure, especially through alpha-helices, leads to fragmentation into clusters of those cycle lengths that are associated with its formation. Intuitively, this process can be understood as {\it collapse} into protein (secondary) structure in the abstract space of cycle length frequencies. From this point of view, we may argue that after the two phase transitions are completed protein structure---when expressed in terms of cycle lengths--- is fully obtained. The least geometrical threshold for which this situation occurs is at about $t\approx 9$ (see Fig. \ref{fig1}), because it right there where the convex region in $H_c(t)$ ends. Interestingly, in previous studies on protein contact maps this value turned out to be special. For example, in \citep{veku1997} it is argued that a choice of $t\approx9$ is determined by the requirement that the average number of C$_\alpha$--C$_\alpha$ contacts for each amino acid is roughly equal to the respective numbers obtained with the all-atom definition of contacts. In other words, for this choice the statistical properties of contact graphs do not depend on the details of the definition of a contact pair. This argument makes well sense in our setting because it is known from  statistical physics that  near a critical point, i.e. near the point of the phase transition, many system properties do not depend on a detailed description of the system itself.  Also in \citep{veku1997} this particular value of $t$ was used to show that a protein 3D structure can be recovered completely from the knowledge of the contact map. This study confirms our previous argument saying that at this value of the threshold parameter the structure is encoded in the contact graph.\smallskip

\begin{figure}[t]
\epsfig{file=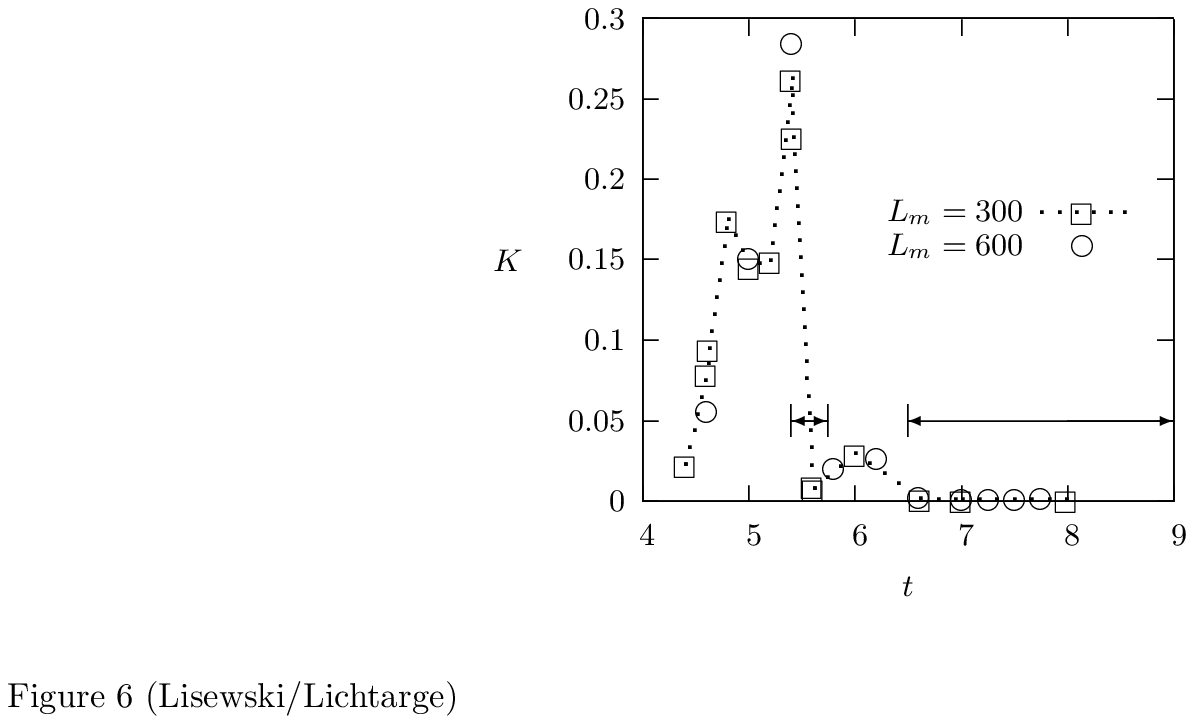}
\caption{The total kinetic energy $K$ as function of the threshold $t$ realized over two choices ($L_m = 300$ and $L_m = 600$) of statistical ensembles. Strong fluctuations in $K$ are the precursors of two first order phase transitions identified by convex regions in $H_c(t)$, see Fig. \ref{fig1}, and the two horizontal arrows above assign those two convex regions. \label{fig6}}
\end{figure}

Beside convex regions in the entropy function, phase transitions in small systems are known to be accompanied by large fluctuations in kinetic energy \citep{guch2000, lara1998}, and we want to verify this fact in our approach. Using again the DALI/FSSP database, Fig. \ref{fig6} shows the total kinetic energy $K(t)$. We observe strong fluctuations of $K$ for $t$-values smaller than  $t \approx 5.5$, that is right before the first order phase transition located at $t_* \approx 5.6$ (see Fig. \ref{fig1}). These fluctuations sharply drop for slightly larger values of $t$. There is a second peak of smaller amplitude in $K$ at $t \approx 6$ indicating the onset of a second first order phase transition at the second convex region in $H_c(t)$. The identification of the emerged system fragments is done by selecting those groups of particles which show the largest fluctuations in kinetic energy; in agreement with our earlier arguments we find in both transitions the main fragment consisting of the particles labeled with 3, 4, and 5. The observed  fluctuations in kinetic energy $K$ become an important validation of our proposed theoretical model because they  show that the constructed Hamilton function $(K+U)$ indeed is a natural choice to be an indicator for the transitory behavior between phases. That is to say, negatively, that an {\it unnatural} choice of $K$ would not be a specific indicator for the phase transitions considered.\smallskip

As the last point of our model analysis we want to demonstrate that the mechanical system is one-dimensional. In \citep{gr2001b} it is argued that the depth $\Delta s$ of a convex intruder in the entropy per particle $s = S/N$ scales with the number of surface particles at the spatial phase boundaries, thus it is
$
\Delta s \sim N^{-1/d} \,,
$  
with $d$ being the spatial dimensionality. For our chosen ensemble $\pr$ we have calculated the entropy depth of the convex intruder $\Delta H_c$ in the range of maximum chain length $ 100 \leq L_m \leq 1058$ (see also Fig. \ref{fig1} for the construction of $\Delta H_c$). Within this range we have not found a significant change of $\Delta H_c$, and we therefore obtain $\Delta s \sim (n^+)^{-1}$. Since the number of surface particles is independent of the total number of particles only in systems with dimensionality one, we conclude that $d = 1$ for the finite system of particles represented by the Hamilton equations (\ref{hamil1}) and (\ref{hamil2}).  Our dimensionality argument signals that the system {\it essentially} behaves like multi-particle system in one dimension. By this term we mean that we cannot arbitrary remove particles and concurrently expect that the transitory behavior between phases remains. For example, by removing the particles labeled with $k \in \{3,4,5\}$ the system would not exhibit a phase transition at $t_*$ anymore, and consequently $\Delta s$ would cease to have a physical meaning. But as long as this phase transition is preserved, the corresponding depth of the convex intruder is essentially independent from the particle number---thus suggesting one-dimensionality.\vspace{-.3cm}

\section{Conclusions and outlook\label{conclusion}}

We have described several unsuspected features of protein folds by introducing a representation based on simple statistical observables. They are defined as frequencies of cycles of different integer length $k$ in contact graphs, and we have shown that they encode a characteristic statistical signature when expressed by their Shannon entropy $H_c$. This signature consists of convex regions in the entropy when considered as a function of the geometrical contact threshold $t$. Based on this representation, we have proposed a classical Hamiltonian system as a mechanical model for the observed behavior of the entropy function $H_c(t)$. This model shows several striking features including chirality, broken time reversal symmetry, first order phase transitions in the context of small statistical systems, and one-dimensionality. Concurrently, it allows to see convex regions in the entropy function as regions where fragmentation of the mechanical system constituents occurs in a first-order phase transition reflecting the generation of protein structure. This transition-- also known as (multi-)fragmentation transition in other statistical systems far from a thermodynamic limit-- has for the first time been directly studied for protein structures.\smallskip

In a broader sense, our approach offers an extended view on protein structures. Although one is still free to consider them as {\it analytically} deduced-- that is, derived from an a priori given ensemble $\pr$ of real protein data such as the Protein Data Bank-- there is now an alternative mode of description in which those observables, and within them the encoded information about protein structure,  are obtained {\it synthetically} through the process of actually solving the Hamiltonian system once its potential energy and the initial conditions have been properly identified. Of course, such an synthetic approach would exist concurrently to known theoretical models and computational algorithms that describe the behavior of protein folds. It is worthwhile to stress, though, that our proposed mechanical system has significantly fewer degrees of freedom than does a (quantum) physical set up necessary to describe protein dynamics in usual 3D space, and hence its direct computational study would require much lower complexity. Thus we think that forthcoming studies of the proposed mechanical system should look to identify the potential energy function in order to equip our model with true predictive power.\smallskip

In the context of the protein design problem, this representation of protein structures is a specific realization of the protein design alphabet. This proposition has found strong support in our results showing that in native protein structures the amino acid usage and the usage of the cycle length alphabet--- the latter defined entirely upon structure--- show striking quantitative similarities. We therefore conclude that amino acid sequences directly encode structural information when represented as lengths of oriented cycles in protein contact graphs.\smallskip

Since in this work we consider only statistical distributions of cycle lengths derived from entire protein folds, we have no specific information at the level of individual amino acids, and therefore we cannot make any predictions about functionally relevant protein residues, for example. However, it has been shown recently that contact frequencies are not uniformly distributed along a protein's polypeptide chain and that residues with high closeness values, i.e. those residues which are neighbors to many other residues in the graph topology, do correlate with functional and evolutionary important sites \citep{amsh2004}. And in \citep{poro2005} the concept of global protein design has been employed for individual residues assessing a site-specific distribution of the amino acid alphabet constrained by local structural features. We therefore believe that our results may be reasonably applied on a local, i.e. residue specific, level as well resulting in a study of distributions $\{q^{(i)}_k\}$ at a chosen residue site $i$. Such future applications would bring us closer to established methods such as the evolutionary trace that looks for statistically significant correlations between sequence, structure and function based on evolutionary variation \citep{lich1996, lich2003}. \medskip

We should finally note that there is another example of a quantitative measure being apparently related to the cycle frequencies $q_k$ introduced here. It is referred to as the {\it contact order} ($CO$) originally introduced in \citep{plsi1998} as $CO=1/(n L) \sum_{j>i} d(i,j)$, where $n$ is the number of contacts in a protein and $d(i,j)$ is the sequence distance function between residues $i$ and $j$. This measure has been mainly studied in the context of the protein folding process. On the contrary, our analysis focuses on already folded native protein structures and although sequence separation between residues is used in our definition of $q_k$ as well, $CO$ is a different mathematical construction from any of those we have established in this work. It is therefore fair to say that previous studies based on the contact order are not directly related to our results.\medskip 

The authors thank I. Mihalek and I. Re\v{s} for their helpful comments. Financial support from the grants NIH R01-GM66099, NSF DBI-0318415 and from the Keck Foundation is gratefully acknowledged.

{\small \bibliography{text+figures-aml}}

\pagebreak
\pagestyle{empty}
\vfill

\vfill

\vfill

\vfill

\vfill

\vfill

\end{document}